\documentclass[conference]{IEEEtran}
\IEEEoverridecommandlockouts
\usepackage{cite}
\usepackage{amsmath,amssymb,amsfonts}
\usepackage{algorithmic}
\usepackage{graphicx}
\usepackage{textcomp}
\usepackage{xcolor}
\usepackage{threeparttable}
\usepackage{mathtools}
\usepackage{xspace}
\usepackage{graphicx}
\usepackage{wrapfig}
\usepackage{lscape}
\usepackage{rotating}
\usepackage{epstopdf}
\usepackage{siunitx}
\usepackage{caption}
\usepackage{subcaption}
\usepackage{xspace} 
\usepackage{lipsum}
\usepackage[utf8]{inputenc}
\usepackage{mathtools}
\usepackage{longtable}
\usepackage{booktabs}
\usepackage{algorithm}

\def\BibTeX{{\rm B\kern-.05em{\sc i\kern-.025em b}\kern-.08em
    T\kern-.1667em\lower.7ex\hbox{E}\kern-.125emX}}

\usepackage{fancyhdr}   

\fancypagestyle{firstpage}{     
  \fancyhf{}
  \chead{\textcolor{gray}{This article has been accepted for publication in the 8th IEEE International Conference on Activity and Behavior Computing}}
  \fancyfoot[C]{\small{\textcolor{gray}{~\copyright~ Personal use of this material is permitted.  Permission from IEEE must be obtained for all other uses, in any current or future media, including reprinting/republishing this material for advertising or promotional purposes, creating new collective works, for resale or redistribution to servers \\ or lists, or reuse of any copyrighted component of this work in other works.}}}

}

\begin{document}

\title{Calibration-Free Induced Magnetic Field Indoor and Outdoor Positioning via Data-Driven Modeling}


\author{\IEEEauthorblockN{Qiushi Guo, Matthias Tschöpe, Mengxi Liu, Sizhen Bian, Paul Lukowicz}
\IEEEauthorblockA{\textit{DFKI} \\
Kaiserslautern, Germany \\
name.surname@dfki.de}
}


\maketitle

\begin{abstract}

Induced magnetic field (IMF)–based localization offers a robust alternative to wave-based positioning technologies due to its immunity to non-line-of-sight conditions, environmental dynamics, and wireless interference. However, existing magnetic localization systems typically rely on analytical field inversion, manual calibration, or environment-specific fingerprinting, limiting their scalability and transferability.
This paper presents a calibration-free, data-driven IMF localization framework that directly maps induced magnetic field measurements to spatial coordinates using supervised learning. By replacing explicit field modeling with learning-based inference, the proposed approach captures nonlinear field interactions and environmental effects without requiring environment-specific calibration. An orientation-invariant feature representation enables rotation-independent deployment.
The system is evaluated across multiple indoor environments and an outdoor deployment. Benchmarking across classical and deep learning models shows that a Random Forest regressor achieves sub-20~cm accuracy in 2D and sub-30~cm in 3D localization. Cross-environment validation demonstrates that models trained indoors generalize to outdoor environments without retraining. We further analyze scalability by varying transmitter spacing, showing that coverage and accuracy can be balanced through deployment density. Overall, this work establishes data-driven IMF localization as a scalable and transferable solution for real-world positioning.

\end{abstract}

\begin{IEEEkeywords}
Indoor localization, induced magnetic field, machine learning, regression, neural networks, signal modeling

\end{IEEEkeywords}

\section{Introduction}

\thispagestyle{firstpage} 


Accurate and reliable indoor localization remains a fundamental challenge in ubiquitous computing, robotics, and context-aware systems \cite{garcia2024relabeling, zafari2019survey, hamdhana2024summary, mostafa2025survey, lago2019nurse,leitch2023indoor}. A wide range of technologies has been explored across both academic research and commercial applications, including ultrasound \cite{crețu2022evaluation, hoeflinger2019passive, sainjeon2016real,famili2022rail}, Wi-Fi/BLE fingerprinting \cite{zhang2023autoloc,garcia2024arelabeling, song2019novel,roy2022survey}, image processing \cite{xie2022deep,alam2023review,heya2018image, morar2020comprehensive}, radio-frequency identification (RFID) \cite{ali2024low, bardareh2022integrated,seco2018smartphone, fang2016case}, and inertial sensing \cite{dyhdalovych2025particle,jiang2024robust, sun2022indoor, poulose2019hybrid, poulose2019indoor}. While these approaches offer distinct advantages, each is subject to domain-specific limitations. Optical and ultrasonic systems typically require line-of-sight (LOS) conditions and are sensitive to environmental changes \cite{hammoud2016robust, rahman2020recent}. Radio-based techniques such as Wi-Fi and BLE suffer from multipath interference and environmental dynamics \cite{leitch2023indoor,bahle2021using}, whereas inertial methods accumulate drift over time \cite{ornhag2022trust}.

In contrast to wave-based localization, magnetic field–based localization operates in a fundamentally different physical regime. Low-frequency magnetic fields exhibit quasi-static near-field behavior and are largely immune to non-line-of-sight (NLOS) conditions, dynamic obstacles, illumination changes, and acoustic noise \cite{bian2020wearable, ouyang2022survey, 11118455, shirai2019dc,bian2020social}. Magnetic fields propagate through most non-metallic materials with minimal attenuation, making them well suited for cluttered, visually obstructed, or dynamically changing environments. Moreover, magnetic sensing is independent of wireless interference and ambient lighting—factors that frequently degrade the performance of wave-based systems. These properties make magnetic localization particularly attractive in challenging scenarios such as industrial spaces, dense indoor environments, and even underwater or subterranean settings, where optical and RF signals attenuate severely \cite{ren2025optimizing, bian2021induced, bian2025magnetic,bian20233d}. Table~\ref{tab:Localization Technology Comparison} summarizes the representative advantages and limitations of commonly used localization technologies.

Despite these advantages, existing magnetic localization systems remain constrained by their reliance on analytical field inversion, manual calibration, or environment-specific fingerprinting \cite{bian2022human}. Analytical dipole-based models struggle to capture nonlinear field distortions caused by metallic structures, irregular coil geometries, or deployment variability, often requiring recalibration when the environment changes. As a result, scalability and transferability across environments remain open challenges.

To address these limitations, this work introduces a \emph{learning-based induced magnetic field (IMF) localization paradigm} that directly maps induced magnetic field measurements to spatial coordinates using supervised machine learning. Unlike prior approaches that explicitly invert magnetic field equations or depend on handcrafted fingerprints, the proposed framework implicitly learns nonlinear field–position relationships from data. This shift enables \textbf{calibration-free deployment}, \textbf{orientation-invariant localization}, and \textbf{cross-environment generalization} without requiring environment-specific tuning.

Finally, while magnetic field strength inherently decays with distance according to physical laws, we treat scalability as a \emph{system-level design trade-off} rather than a fundamental limitation. By systematically studying transmitter spacing and deployment geometry, we demonstrate how localization accuracy and coverage can be balanced through transmitter density, enabling practical scalability for real-world applications. Together, these contributions position induced magnetic field sensing not merely as an alternative localization modality, but as a robust, learning-enabled localization infrastructure capable of stable performance across diverse indoor and outdoor environments.

\subsection{Related Work in Magnetic Field-Based Localization}

Magnetic field-based localization has emerged as an attractive alternative to traditional RF or vision-based systems due to its immunity to non-line-of-sight (NLOS) effects and reduced sensitivity to environmental interference. Existing research spans theoretical modeling, system design, and practical deployment.

Pirkl \textit{et al.} \cite{pirkl2012robust} developed an indoor positioning system based on resonant magnetic coupling, achieving sub-meter accuracy while maintaining robustness against dynamic environmental changes such as human movement. This work demonstrated the feasibility of low-frequency magnetic fields for stable and interference-resistant indoor localization. Building on this foundation, Solin \textit{et al.} \cite{solin2018modeling} proposed a Bayesian nonparametric framework using Gaussian processes to model and interpolate ambient magnetic fields, enabling continuous localization while accounting for spatial variations and temporal drift.

From an analytical perspective, Pasku \textit{et al.} \cite{pasku2016magnetic} analyzed quasi-stationary magnetic fields generated by coils, comparing practical field distributions with ideal dipole models to derive insights into near-field behavior and 3D positioning accuracy. Chen \textit{et al.} \cite{chen2015integrated} combined Wi-Fi fingerprinting, magnetic sensing, and pedestrian dead reckoning using an Unscented Kalman Filter, improving robustness in complex indoor environments. More recently, low-frequency AC magnetic field localization systems \cite{pasku2016magnetic} have demonstrated sub-30~cm accuracy in both indoor and outdoor testbeds, highlighting the potential scalability of magnetic approaches.

Despite these advances, most existing systems rely on analytical modeling or empirically calibrated fingerprints, which struggle to capture nonlinear field interactions arising from metallic disturbances, irregular coil geometries, or deployment variability. Furthermore, many approaches remain scene-dependent and require recalibration when transferred to new environments. These limitations motivate the need for a data-driven magnetic localization framework capable of learning complex field–position mappings and generalizing across spatial configurations.

\subsection{Contributions}

This work addresses the aforementioned challenges by introducing a comprehensive, machine learning–driven framework for induced magnetic field (IMF) localization. Our contributions advance the state of the art in both system design and experimental validation:

\begin{itemize}
    \item \textbf{A calibration-free, data-driven IMF localization framework:} 
    We depart from analytical and fingerprint-based magnetic localization by introducing a fully data-driven learning paradigm that directly maps induced magnetic field measurements to spatial coordinates, eliminating the need for explicit field inversion or environment-specific calibration.

    \item \textbf{Redesigned magnetic sensing architecture:} 
    The proposed system integrates multiple orthogonal transmitter coils with a tri-axial receiver to enable complete spatial field capture. The hardware is optimized for stable field generation and reliable sensing across diverse deployment conditions.

    \item \textbf{Orientation-invariant feature representation:} 
    We introduce an orientation-invariant feature formulation that mitigates the effects of receiver rotation and sensor misalignment, enabling rotation-independent localization and improving cross-scene generalization.

    \item \textbf{Systematic benchmarking of learning models:} 
    We evaluate classical regression, ensemble learning, and deep neural network models to quantify their ability to capture nonlinear magnetic field–position relationships, revealing accuracy–complexity trade-offs relevant for real-time localization.

    \item \textbf{Cross-environment generalization and scalability analysis:} 
    We demonstrate that models trained in one environment can be deployed in structurally different indoor and outdoor environments without retraining, and analyze scalability as a controllable trade-off via transmitter spacing and deployment density.

    \item \textbf{Comprehensive experimental validation:} 
    Extensive experiments across multiple indoor scenes and an outdoor environment show that the proposed approach achieves sub-20~cm accuracy in 2D localization and sub-30~cm accuracy in 3D positioning, outperforming analytical and manually calibrated baselines.
\end{itemize}

In summary, this work bridges analytical magnetic field modeling and learning-based localization, establishing a scalable, transferable, and calibration-free framework for induced magnetic field–based positioning in real-world environments.

\begin{table*}[ht]
\centering
\begin{threeparttable}
\begin{tabular}{p{1.6cm} p{1.5cm} p{6.8cm} p{5.3cm}}
\toprule
    Main Technology & Category\cite{bian2022state} & Advantages & Disadvantages \\
\midrule
    IMU  \cite{poulose2019hybrid, poulose2019indoor} & Mechanical & Self-contained; robust to environmental conditions; continuous localization updates. & Drift inherent to sensors; requires initialization and calibration; provides only relative positioning. \\
\midrule
    Ultrasonic  \cite{hoeflinger2019passive, sainjeon2016real} & Wave & High accuracy under LOS; simple implementation. & Requires synchronization; affected by ambient noise and NLOS conditions. \\
\midrule
    RFID \cite{seco2018smartphone, fang2016case} & Wave & Low cost; passive tags require no power; easy integration with IoT infrastructure. & Limited range and accuracy; active tags require batteries and are more expensive. \\
\midrule
    Radar \cite{sesyuk2024radar, antonucci2019performance} & Wave & High accuracy and penetration. & Requires LOS; sensitive to reflections and multipath effects. \\
\midrule
    WIFI/BLE \cite{garcia2024arelabeling, song2019novel} & Wave & Utilizes existing infrastructure; low additional cost. & Coarse localization; sensitive to interference and environmental dynamics; often requires calibration. \\
\midrule
    Induced Magnetic Field & Field & Immune to NLOS and illumination changes; robust to moving obstacles; minimal interference; effective in cluttered and underwater environments. & Limited operational range; accuracy affected by metallic structures; requires coil calibration and field modeling. \\
\bottomrule
\end{tabular}
\caption{Comparison of representative indoor localization technologies. Adapted from Jorge \cite{torres2010review}.}
\label{tab:Localization Technology Comparison}
\end{threeparttable}
\end{table*}

\section{Hardware Implementation}

\subsection{Physical Background}
Artificially generated magnetic field-based positioning systems rely on the well-defined relationship between magnetic field strength and the relative distance between transmitter and receiver coils. In our system, a low-frequency alternating current (AC) magnetic field is generated using multiple coils. Assuming the coordinate origin is located at the coil center and the $z$-axis aligns with the coil’s normal, the magnetic flux density $B$ along the $z$-axis can be expressed in a simplified dipole form \cite{jackson2007classical}:
\begin{equation}
B(x, y, z, t) = \frac{\mu_0}{4\pi} \left[ 3\vec{d}(\vec{m}\cdot\vec{d}) - \vec{m}d^2 \right] d^{-5} e^{-j\omega t}
\end{equation}
where the magnetic dipole moment $\vec{m}$ is defined as
\begin{equation}
\vec{m} = NIR^2\vec{n}
\end{equation}
with $N$ denoting the number of turns, $I$ the peak current, $R$ the coil radius, and $\vec{n}$ the unit vector normal to the coil plane. The remaining parameters are defined as follows:
\begin{itemize}
    \item $x, y, z$: spatial coordinates of the observation point;
    \item $\mu_0$: magnetic permeability of free space;
    \item $\omega$: angular frequency of the excitation current;
    \item $\vec{d}$: vector from the transmitter to the observation point;
    \item $d = |\vec{d}|$: distance between transmitter and observation point.
\end{itemize}

To model the relationship between distance and induced voltage, the magnetic potential can be approximated as \cite{pasku2016magnetic}:
\begin{equation}
V_p = \frac{\mu_0 m N \omega}{4\pi} f_{\text{approx}}(x, y, z)
\end{equation}
where
\begin{equation}
f_{\text{approx}}(x, y, z) = S\frac{-x^2 - y^2 - 2z^2}{(x^2 + y^2 + z^2)^{2/5}}
\end{equation}
serves as an analytical simplification of the full magnetic field expression:
\begin{equation}
f(x, y, z) = \iint \frac{-h^2\rho - \rho^3 - 2\alpha\rho^2 + 2\rho z^2}{(h^2 + 2\rho\alpha + \rho^2 + z^2)^{2/5}}\, d\theta\, d\rho
\end{equation}
This approximation eliminates computationally expensive double integration, enabling real-time field sampling and localization in practical systems.

\begin{figure}[ht]
    \centering
    \includegraphics[width=0.9\linewidth]{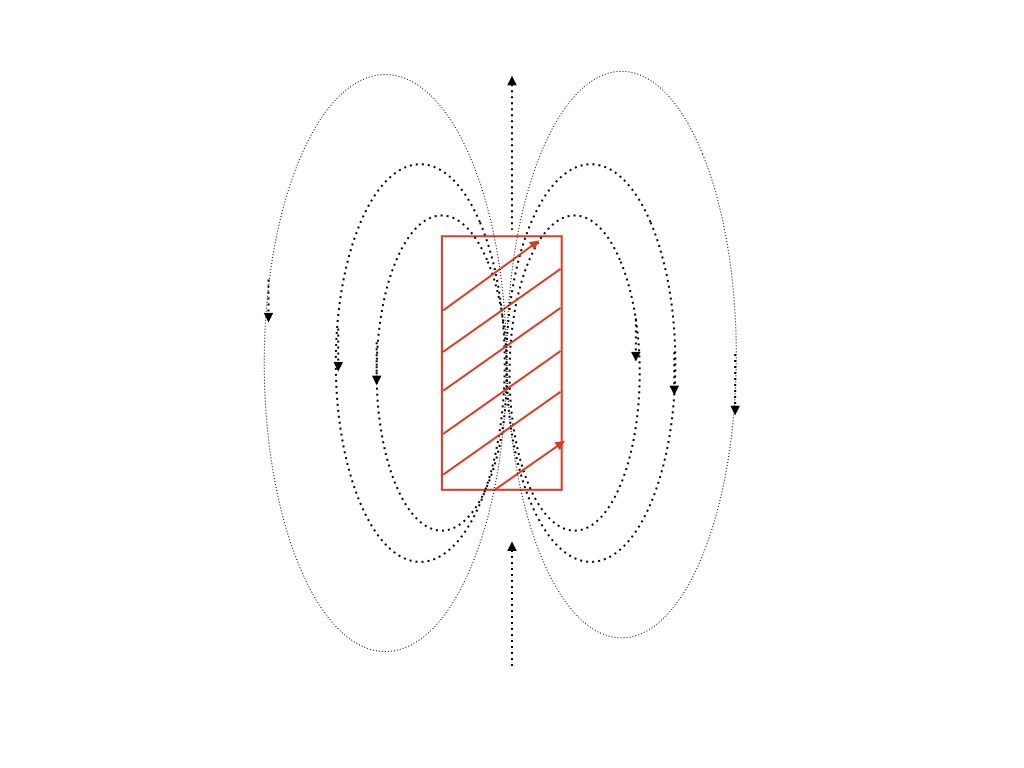}
    \caption{Transmitter coil prototype for low-frequency magnetic field generation.}
    \label{fig:coil}
\end{figure}

Our system employs five transmitter coils that generate oscillating magnetic fields at predefined low frequencies. Each transmitter acts as a signal source, while the receiver—an oscillating LC circuit tuned to the same frequency—detects the induced voltage. As illustrated in Fig.~\ref{fig:coil}, the induced voltage amplitude directly correlates with the magnetic field strength and inversely with the cube of the transmitter–receiver distance.

At longer ranges, the magnetic field magnitude decays proportionally to $1/d^3$. The proposed localization framework is based on \textit{resonant magnetic coupling}, a well-studied principle in wireless power transfer and near-field communication. According to Faraday’s law of induction, the voltage induced in a receiver coil is proportional to the time derivative of the magnetic flux it encloses. Consequently, the induced voltage depends on the transmitter current amplitude and frequency, and inversely on the cube of the separation distance.

\begin{figure}[ht]
    \centering
    \includegraphics[width=0.7\linewidth]{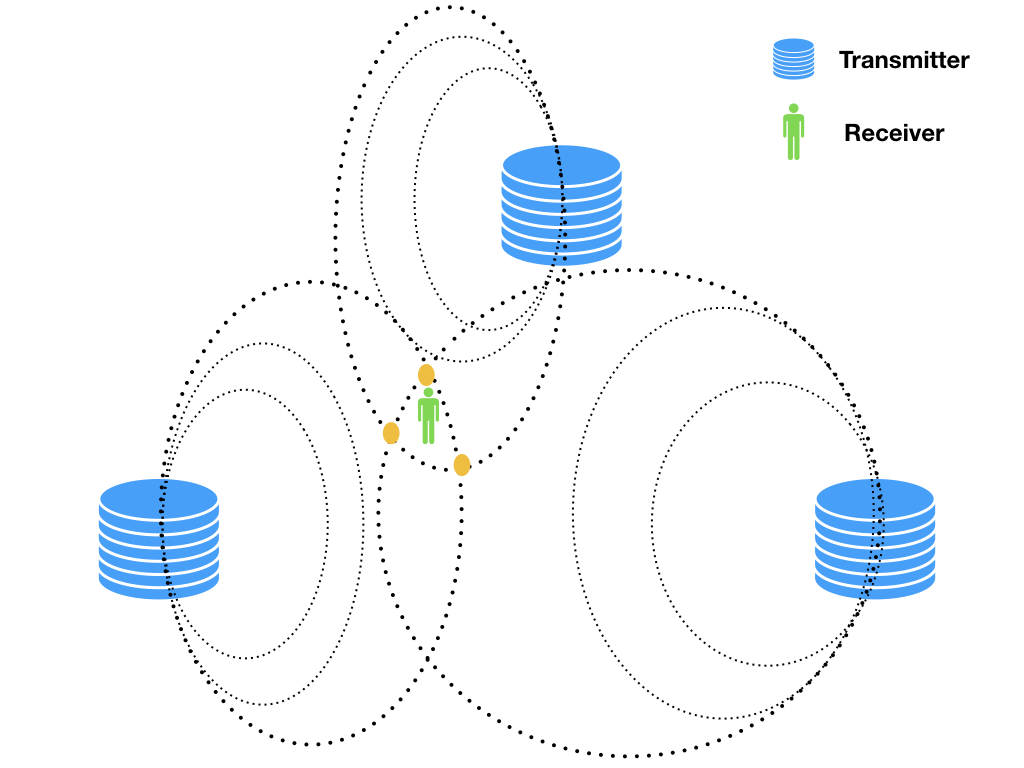}
    \caption{Geometric principle of 3D localization using multiple magnetic field transmitters.}
    \label{fig:location}
\end{figure}

In a simplified conceptual model, the magnetic field generated by each coil can be visualized as a set of concentric isosurfaces of constant field strength. In 3D space, the intersection of two such surfaces forms a circle; therefore, at least three transmitters are required to uniquely determine a point’s position. Figure~\ref{fig:location} illustrates this geometric principle. In realistic environments containing metallic objects or magnetic materials, field distortion occurs, complicating analytical inversion—further motivating our data-driven learning approach.

\subsection{Hardware Architecture}
The complete hardware system consists of two main subsystems: a multi-axis transmitter array and a tri-axial receiver module. Each transmitter includes three orthogonal coils that generate alternating magnetic fields sequentially at 20~kHz. The receiver, likewise, comprises three orthogonal sensing coils for measuring induced voltages in the $x$, $y$, and $z$ directions.

The analog front-end processes the weak induced voltages through filtering, amplification, and analog-to-digital conversion. A fourth-order Butterworth bandpass filter centered at 20~kHz provides both gain and frequency selectivity, ensuring suppression of out-of-band noise. The Butterworth topology was selected for its maximally flat amplitude response and predictable phase behavior, making it ideal for anti-aliasing before digitization. The filtered signal is then fed into a logarithmic amplifier that compresses its wide dynamic range into a decibel scale via nonlinear transformation. This step stabilizes amplitude variations and improves measurement linearity.

Compared to previous designs, the updated circuit exhibits significantly reduced noise and higher measurement precision. The modular design enables flexible configuration of the transmitter–receiver geometry, facilitating controlled experiments for magnetic field mapping and learning-based localization.

\section{Experiment}

\subsection{Devices}
\subsubsection{Transmitter}
\begin{figure}[ht]
\centering
\begin{minipage}[t]{0.45\linewidth}
\centering
\includegraphics[width=0.8\textwidth,height=2.5cm]{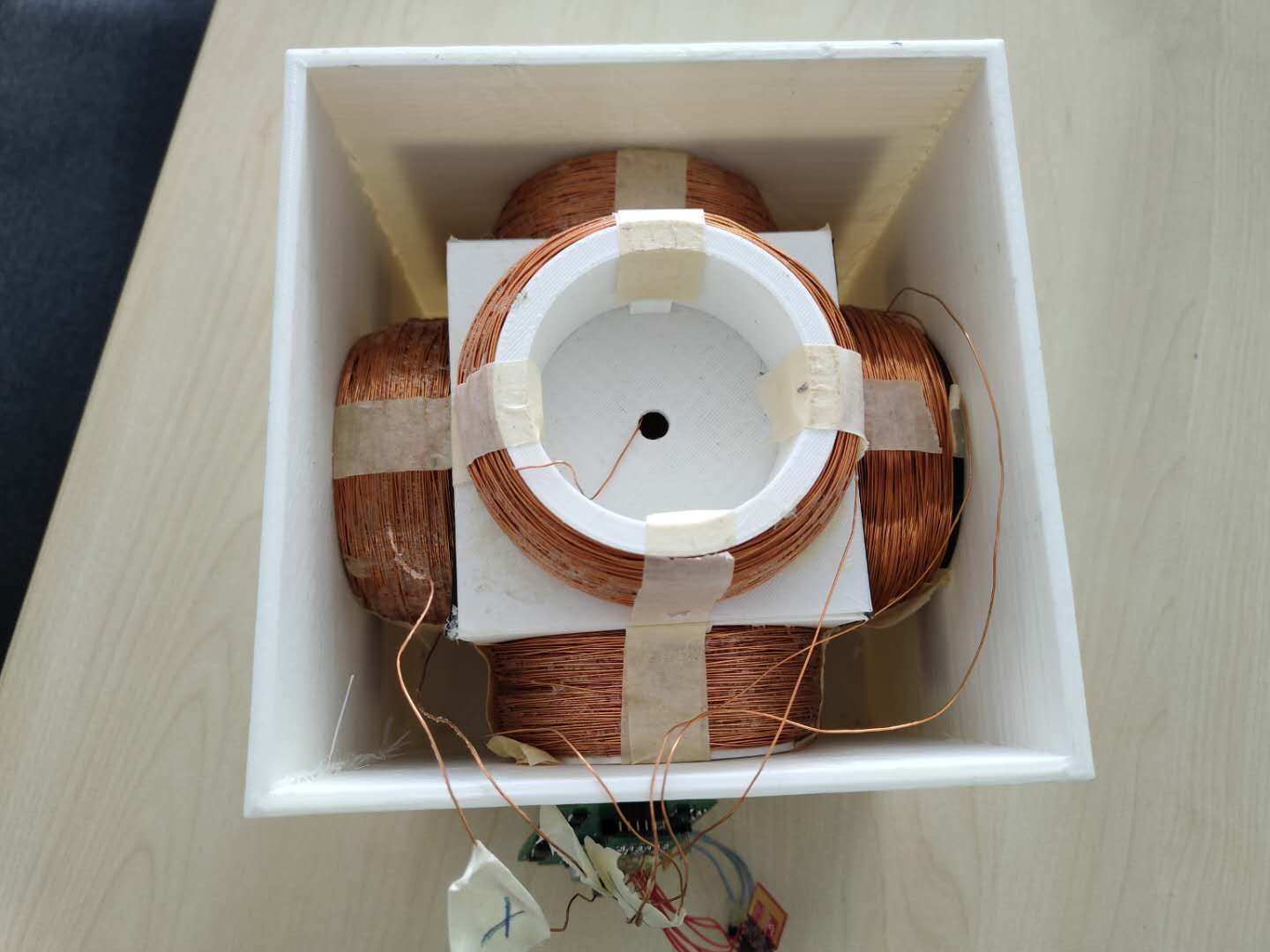}
\end{minipage}
\quad
\begin{minipage}[t]{0.45\linewidth}
\centering
\includegraphics[width=0.8\textwidth,height=2.5cm]{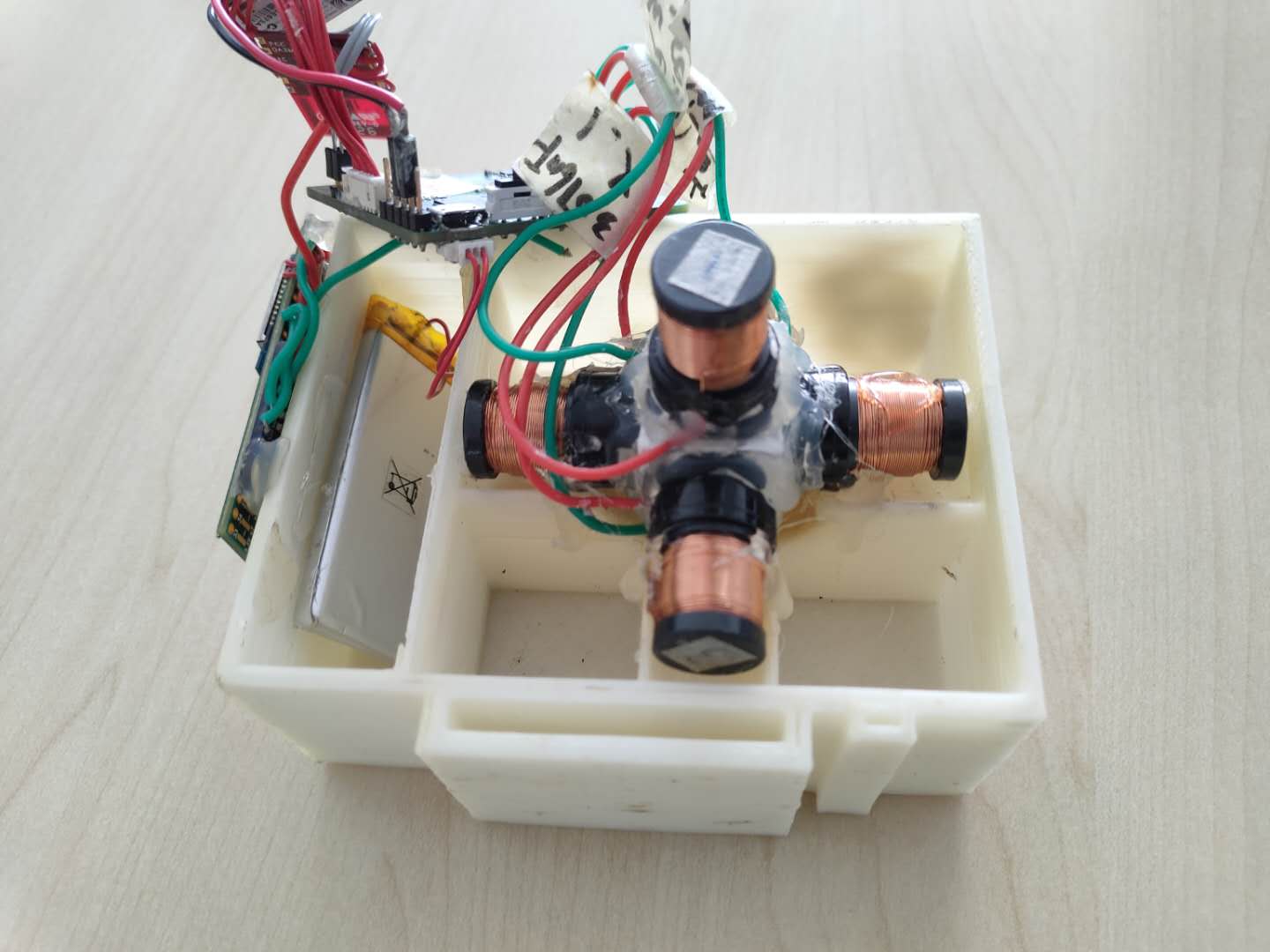}
\end{minipage}
\caption{Prototype of the transmitter (left) and receiver (right).}
\label{Transmitter and Receiver}
\end{figure}

Each transmitter unit consists of three orthogonal coils ($x$, $y$, and $z$ axes), each wound with several hundred turns of copper wire. The coils are housed within a protective plastic enclosure and connected to a custom-designed PCB that synchronizes the activation sequence across multiple transmitters. During operation, the transmitters are sequentially activated in a time-multiplexed manner: at any instant, only one axis from one transmitter is active. For example, with three transmitters deployed, transmitter 1 sequentially excites its $x$-, $y$-, and $z$-axis coils before control passes to transmitter 2, and subsequently transmitter 3. This cyclic excitation pattern ensures clear signal separation across transmitters and axes. An example of the transmitter and receiver prototypes is shown in Fig.~\ref{Transmitter and Receiver}.

\subsubsection{Receiver}
The receiver module also incorporates three perpendicular sensing coils, though with smaller dimensions compared to the transmitter coils. The oscillating magnetic fields generated by the transmitters induce voltages within the receiver coils, which are subsequently filtered and digitized. During data collection, the receiver is moved throughout the measurement space to capture signal strength variations corresponding to different spatial locations. Data collection was performed in an orientation-free manner, ensuring that model learning remains invariant to sensor rotation.

\subsubsection{Ultrasound Ground Truth System}
An ultrasound-based localization system was employed to obtain the ground-truth positions of the receiver during data collection. The system used is the Marvelmind Precise Indoor Navigation System—one of the most accurate commercially available indoor positioning solutions—with a reported precision of approximately 2~cm in ideal conditions.

The system comprises three main components:
\begin{itemize}
    \item A mobile beacon attached to the receiver tray;
    \item Four stationary beacons fixed to the surrounding walls;
    \item A router that synchronizes and records beacon data.
\end{itemize}
Prior to data acquisition, stationary beacons were mounted on walls without obstructions between any pair, ensuring line-of-sight signal propagation. Using the Marvelmind Dashboard software, the stationary beacons were calibrated and their coordinates frozen. Once initialized, the router continuously tracked the 3D position of the mobile beacon.

\begin{figure}[ht]
\centering
\begin{minipage}[t]{0.45\linewidth}
\centering
\includegraphics[width=0.8\textwidth,height=2.5cm]{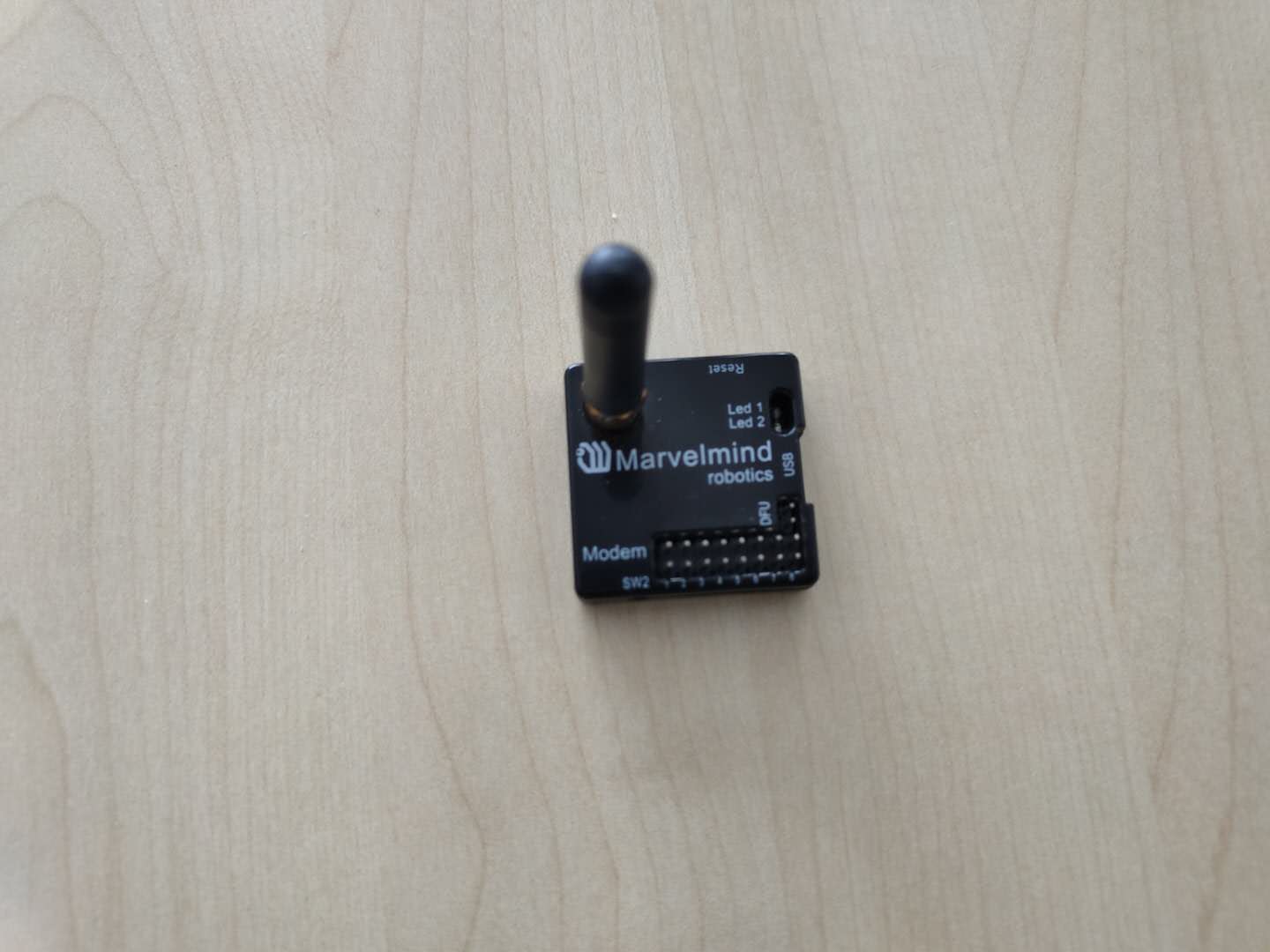}
\end{minipage}
\quad
\begin{minipage}[t]{0.45\linewidth}
\centering
\includegraphics[width=0.8\textwidth,height=2.5cm]{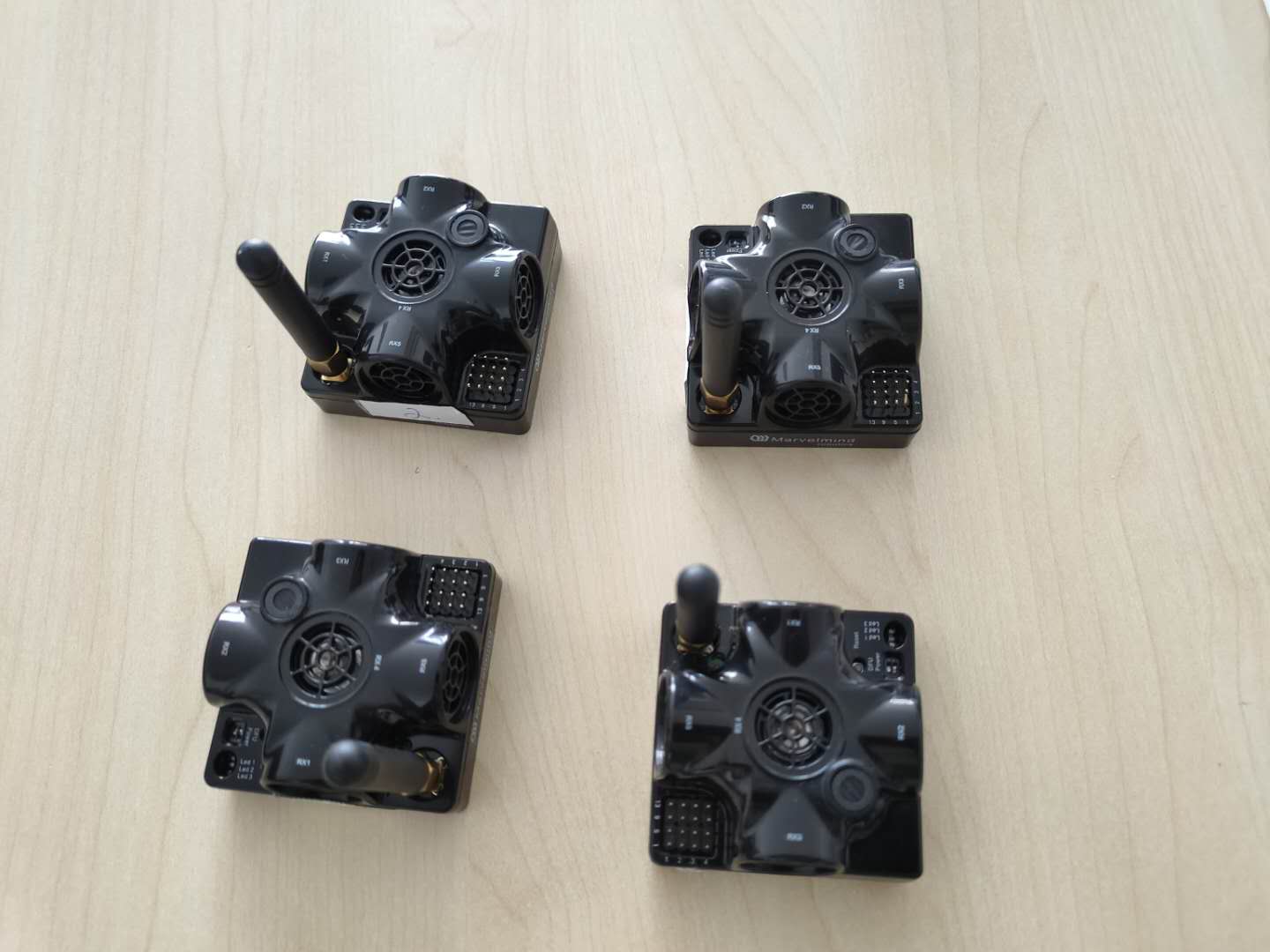}
\end{minipage}
\caption{Marvelmind ultrasonic localization system: router (left) and stationary beacon (right).}
\label{router and ultrasound}
\end{figure}


\subsection{Experimental Environments}
Data collection was conducted across four distinct environments within and around the campus to evaluate the robustness of the proposed localization framework:
\begin{enumerate}
    \item \textbf{Meeting Room:} Located on the intermediate floor of a building, equipped with four stationary beacons placed on walls and windows at a height of 180~cm. 
    \item \textbf{Social Area:} A large open indoor space configured for 3D data collection. Stationary beacons were mounted at 190~cm to extend the operational height range. Experiments were conducted using both three- and five-transmitter configurations.
    \item \textbf{Corridor:} A narrow indoor passageway on the same floor. Four stationary beacons were deployed with clear line-of-sight conditions. Three transmitters were used.
    \item \textbf{Outdoor Area:} Measurements were collected outside the building. Stationary ultrasound beacons were mounted on poles, and five transmitters were placed approximately 4~m apart. 
\end{enumerate}

\begin{figure}[ht]
\centering
\begin{minipage}[t]{0.45\linewidth}
\centering
\includegraphics[width=0.95\textwidth,height=2.5cm]{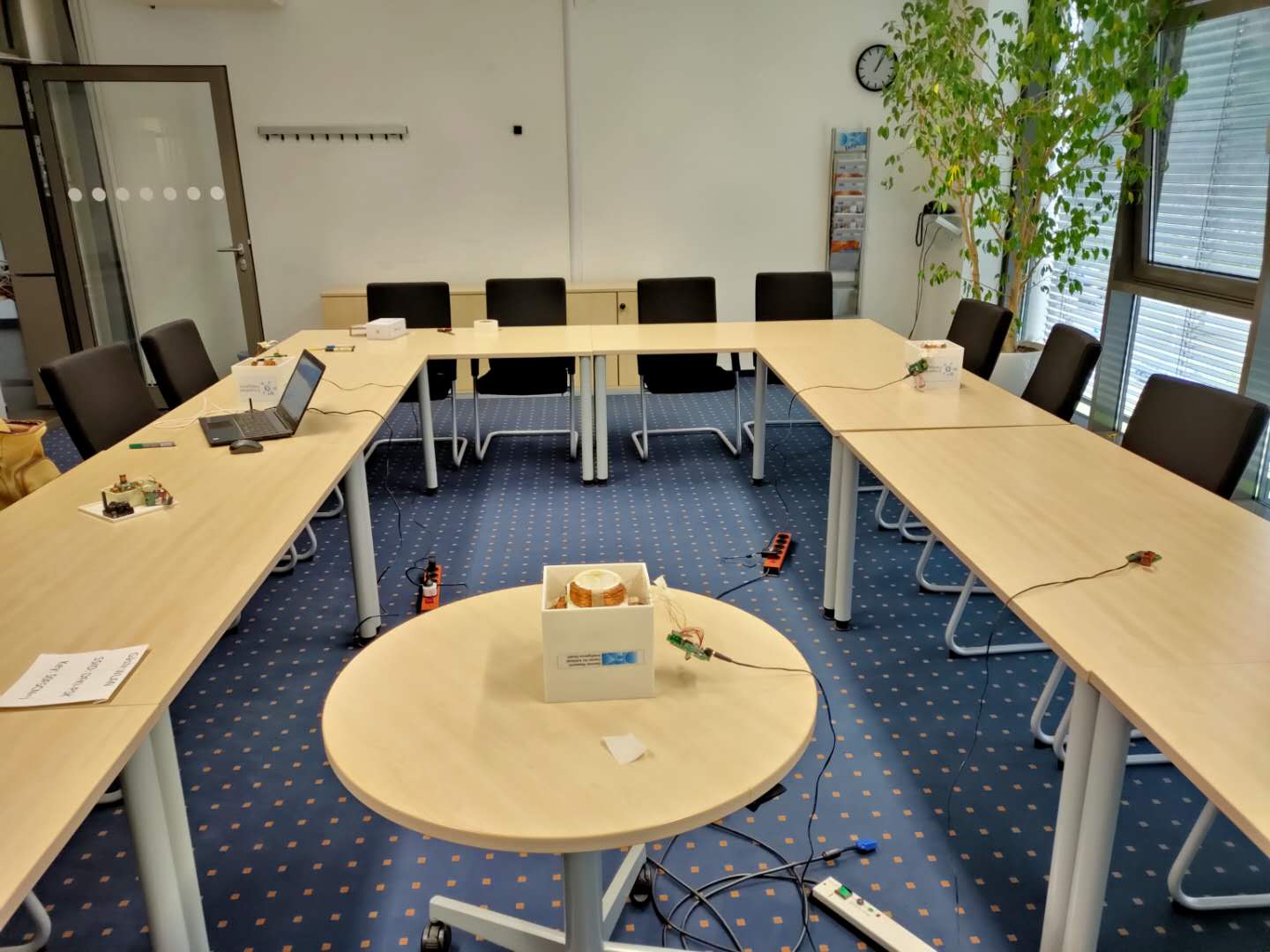}
\end{minipage}
\begin{minipage}[t]{0.45\linewidth}
\centering
\includegraphics[width=0.95\textwidth,height=2.5cm]{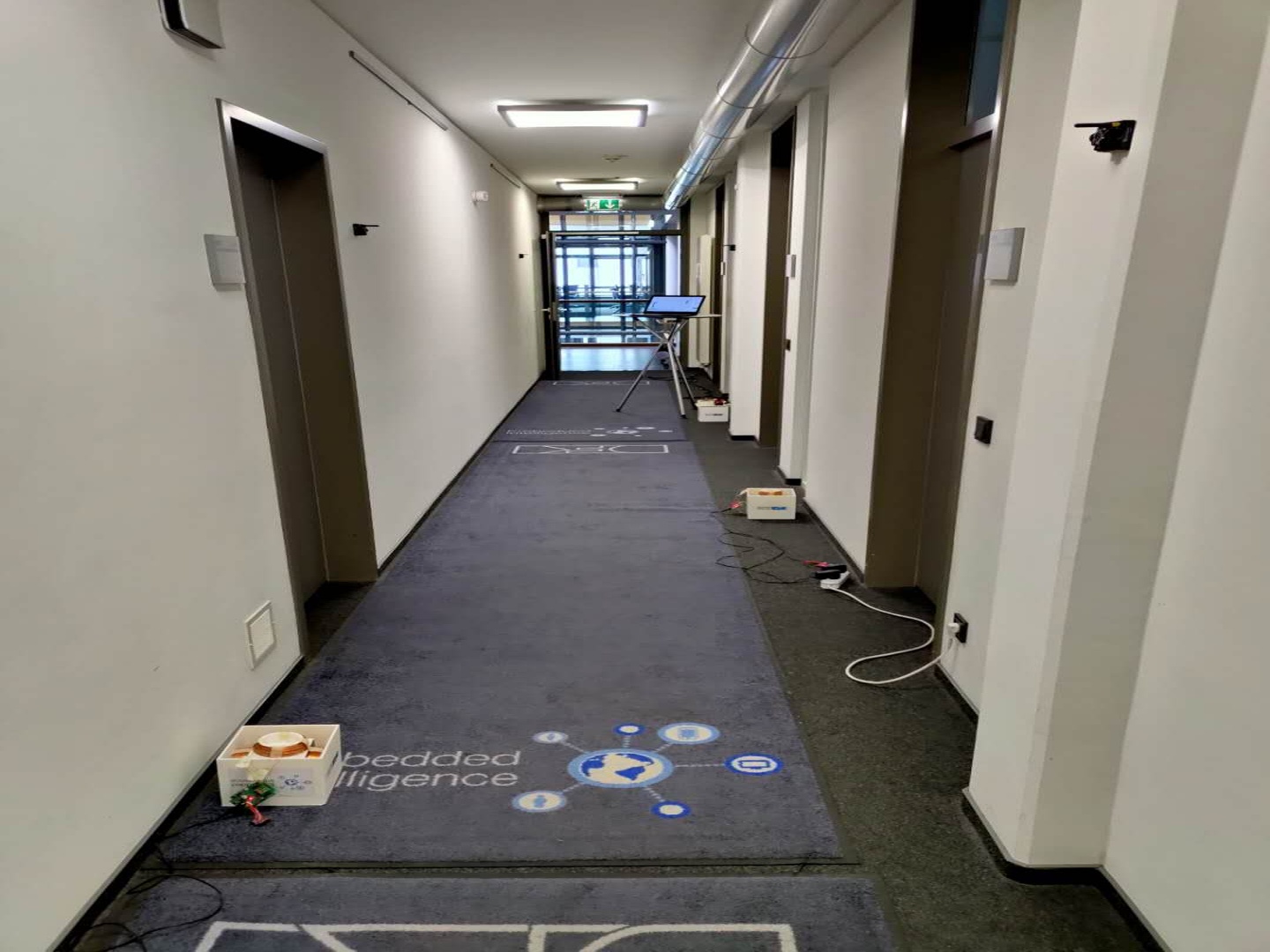}
\end{minipage}
\begin{minipage}[t]{0.45\linewidth}
\centering
\includegraphics[width=0.95\textwidth,height=2.5cm]{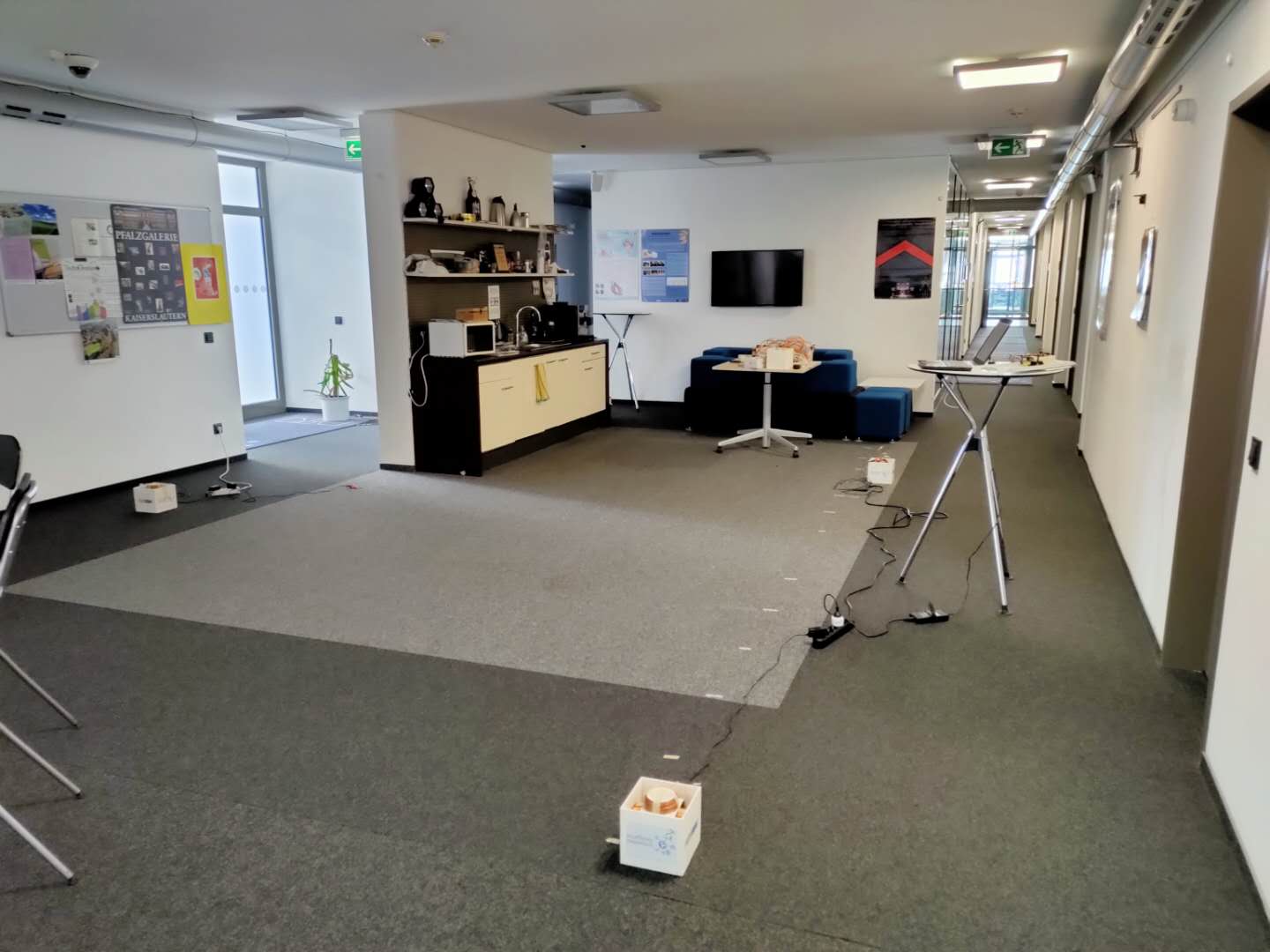}
\end{minipage}
\begin{minipage}[t]{0.45\linewidth}
\centering
\includegraphics[width=0.95\textwidth,height=2.5cm]{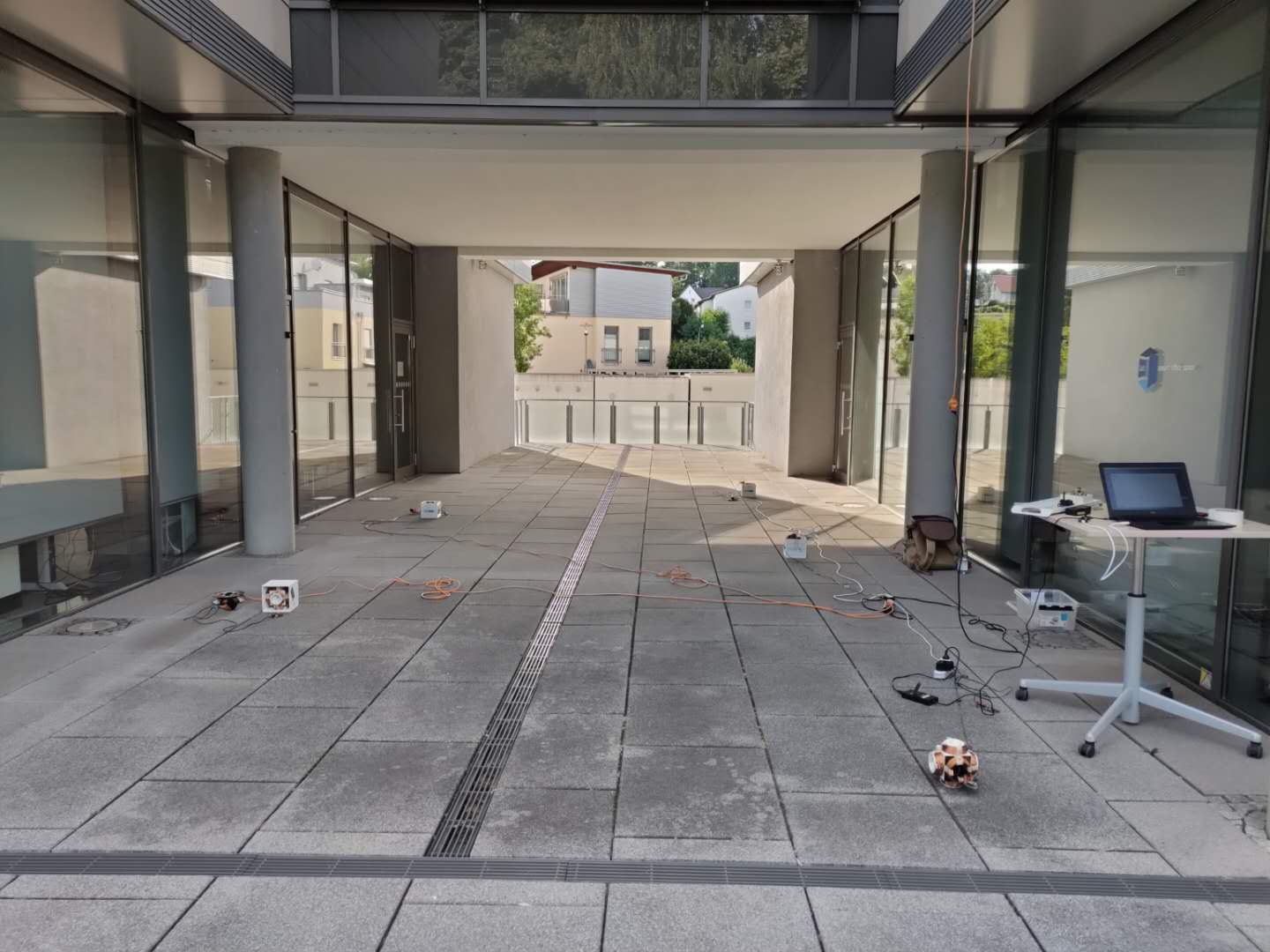}
\end{minipage}
\caption{Experimental scenes for data collection: top-left: Meeting room, top-right: Corridor, bottom-left: Social area, and bottom-right: Outdoor environment.}
\label{scenes}
\end{figure}

This diverse set of environments allows for evaluation of both spatial generalization and environmental robustness, forming the foundation for training and testing the machine learning–based localization models.

\subsection{Training Procedure and Dataset Description}

\subsubsection{Loss Function and Learning Objective}
All models evaluated in this work are trained to minimize the mean squared error (MSE) between the predicted position and the ground-truth position obtained from the ultrasound-based localization system. For a predicted position $\hat{\mathbf{p}}_i$ and corresponding ground truth $\mathbf{p}_i$, the loss is defined as
\begin{equation}
\mathcal{L} = \frac{1}{N} \sum_{i=1}^{N} \|\hat{\mathbf{p}}_i - \mathbf{p}_i\|_2^2,
\end{equation}
where $N$ denotes the number of training samples. For tree-based ensemble methods such as Random Forest, this objective is optimized implicitly during training by minimizing the variance of target values within leaf nodes.

\subsubsection{Data Collection Protocol}
Datasets were collected by keeping the magnetic receiver attached to the ultrasound mobile beacon and moving it freely throughout the measurement space. To ensure diverse spatial coverage, the operator walked randomly within the test area during data acquisition. For each environment and transmitter configuration, data were recorded in three independent sessions, each lasting approximately 30 minutes.

In the three-transmitter setup, one full excitation cycle (covering all transmitter axes) is sampled at approximately 8.3~Hz. In the five-transmitter setup, the sampling rate of a full cycle is approximately 5~Hz, as each transmitter is allocated a 40~ms time window for magnetic field generation. Each full cycle yields one feature vector paired with a ground-truth position.

\subsubsection{Dataset Size and Train--Test Split}
Following the three-session acquisition protocol, the first two sessions were used exclusively for training, while the third session was held out for testing. This session-level split prevents temporal overlap between training and testing data and avoids information leakage across sequences. Using this protocol, approximately 30{,}000 samples were used for training and approximately 15{,}000 samples were used for testing for each experimental configuration.

Unless otherwise stated, all reported results are obtained using this fixed train--test split. No samples from the testing session were used during training or model selection.

\subsubsection{Sequence Handling and Transmitter Configuration}
Although individual samples are treated as independent for model training, sequence-level information is preserved during evaluation for trajectory-based analysis. For sequence experiments, predictions are generated on temporally ordered test samples only, ensuring that no data from the same temporal sequence appear in both training and testing sets.

Transmitter placement and geometry were kept fixed within each experimental configuration. When evaluating cross-environment generalization, identical transmitter layouts were used to avoid introducing unintended domain gaps due to geometry changes. The proposed learning framework is agnostic to the number of transmitters; increasing the number of transmitters extends the feature vector dimensionality accordingly without modifying the training or inference pipeline.

\subsection{Preprocessing}

\subsubsection{Signal Detection}
In the three-transmitter system, each full cycle has a duration of 120~ms, divided into three 40~ms sub-periods corresponding to transmitters $T_1$, $T_2$, and $T_3$. Within each sub-period, the transmitter’s three orthogonal coils ($x$, $y$, $z$) are activated sequentially, generating distinct oscillating magnetic fields. These field components differ in direction and magnitude, allowing each coil to contribute unique spatial information to the received signal.

The raw signals (rectified sensed magnetic field strength) acquired by the receiver contain both magnetic field responses and environmental noise. Typical background noise amplitudes range from $5\times10^7$ to $7\times10^7$, while the induced magnetic signals fall within $8\times10^7$ to $1.5\times10^8$, providing sufficient contrast for reliable detection. A fixed threshold of $7.0\times10^7$ was empirically chosen for both indoor and outdoor experiments to distinguish true signals from noise.

Let $S_{ij}$ denote the induced signal strength generated by the $j$-axis coil of the $i$-th transmitter, where $i \in \{1,2,3\}$ and $j \in \{x,y,z\}$. The activation intervals between coils and transmitters are temporally encoded, allowing each $S_{ij}$ to be uniquely identified. For example, the time difference between activations of $S_{1x}$ and $S_{1y}$ is approximately 6~ms, between $S_{2x}$ and $S_{2y}$ is 6~ms, and between $S_{3x}$ and $S_{3y}$ is about 10~ms.

\subsubsection{Feature Selection}
Following noise suppression and signal segmentation, the next step involves constructing the input feature vector. The processed data for each period can be represented as
\begin{equation}
X' = \{S_{t_1,x,x}, S_{t_1,x,y}, \dots, S_{t_3,z,z}\},
\end{equation}
where $S_{t,i,j}$ denotes the signal strength measured at the receiver coil along axis $j$ when the $i$-th transmitter coil along axis $i$ is active. In total, the raw feature vector for a single period has a dimensionality of 27.

An effective feature representation should satisfy two essential criteria:
\begin{enumerate}
    \item \textbf{Location sensitivity:} Each feature vector should uniquely correspond to a single spatial location, enabling the model to learn a one-to-one mapping between feature space and physical space. This requires managing the inherent field symmetry to prevent ambiguous position estimates.
    \item \textbf{Orientation invariance:} The receiver’s orientation should not affect the model’s ability to infer position. Since rotation alters the directional components of the induced voltage ($S_{t,i,j}$), the final feature set must abstract away rotational effects.
\end{enumerate}

To meet these conditions, we propose the following feature formulation:
\begin{equation}
\begin{array}{l}
    X = \{M_{t_1,x}, M_{t_1,y}, M_{t_1,z}, M_{t_2,x}, M_{t_2,y}, M_{t_2,z}, \\ 
    M_{t_3,x}, M_{t_3,y}, M_{t_3,z}\}
  \end{array}
\end{equation}
where $M_{t_i,\phi}$ denotes the overall magnitude of the signal generated by the $\phi$-axis coil of the $i$-th transmitter, defined as
\begin{equation}
M_{t_i,\phi} = \sqrt{S_{t_i,\phi,x}^2 + S_{t_i,\phi,y}^2 + S_{t_i,\phi,z}^2}.
\end{equation}

While the individual signal components vary with receiver rotation, the magnitude $M_{t_i,\phi}$ remains nearly constant for a fixed spatial position. This transformation effectively normalizes orientation variations, enabling the learning model to generalize across arbitrary receiver orientations without additional calibration.

\subsection{Model Training}

To evaluate the most effective learning strategy for induced magnetic field (IMF) localization, we benchmarked a range of models spanning traditional and modern machine learning paradigms. 


Each model was trained and tested on the collected dataset, and its performance was assessed in terms of four key metrics: accuracy, computational speed, robustness under environmental variation, and model complexity (parameter size). The best-performing method was subsequently integrated into the complete localization pipeline.

\subsubsection{Normalization}

Before model training, all input data were normalized to ensure consistent scaling across features. Many algorithms—particularly those involving distance computations (e.g., $k$-Nearest Neighbors, SVM) or gradient-based optimization (e.g., neural networks)—are sensitive to input magnitude differences. Unnormalized data can lead to unstable learning dynamics, biased distance metrics, and slow convergence.


In this work, min–max normalization was adopted due to its stability across bounded sensor values. Experimental results show that normalization significantly improved performance for MLP and $k$NN, while having a limited effect on ensemble-based methods such as Random Forest.

\subsubsection{Hyperparameter Selection}

Each algorithm was tuned using model-specific hyperparameters, as summarized in Table~\ref{tab:hyperparams}.

\begin{table}[h]
\centering
\begin{tabular}{|c|c|}
\hline
\textbf{Model} & \textbf{Key Parameters} \\
\hline
SVM & Kernel type, bandwidth \\
\hline
Random Forest & Tree depth, number of estimators \\
\hline
$k$NN & Number of neighbors ($k$) \\
\hline
AdaBoost & Learning rate, number of estimators \\
\hline
LSTM & Number of layers, hidden units, time steps \\
\hline
MLP & Number of layers, neurons per layer \\
\hline
\end{tabular}
\caption{Primary hyperparameters tuned for each model.}
\label{tab:hyperparams}
\end{table}

Because the search space for optimal parameters is continuous and often high-dimensional, direct exhaustive optimization is infeasible. 
In this study, we primarily used grid search with moderate intervals to balance computational cost and tuning granularity.

\subsection{Sequence Prediction}

Once the model $F$ was trained for single-point prediction, it was extended to sequential localization for continuous tracking. Given an input feature vector $X_p$ at position $P$, the model predicts the estimated location $\hat{P} = F(X_p)$. 

Building upon this, the position sequence $\hat{P}_i = \{ \hat{p}_{i,1}, \hat{p}_{i,2}, \dots, \hat{p}_{i,n} \}$ is generated from a series of consecutive predictions $P_i = \{ p_{i,1}, p_{i,2}, \dots, p_{i,n} \}$. This formulation leverages temporal coherence—similar to dead reckoning—where neighboring points provide contextual information for smoothing trajectories when the sampling interval is sufficiently short.

To mitigate local anomalies and outliers, a post-processing filter was applied to each predicted sequence. In a typical 10-point trajectory, no more than three points were identified as outliers and subsequently removed using an isolation forest or variance-based adaptive filter (Alg. \ref{alg:Framwork}). The resulting smoothed sequence provides a robust trajectory estimate with improved continuity and spatial stability. 

\begin{figure}[H]
    \centering
    \includegraphics[width=0.8\linewidth]{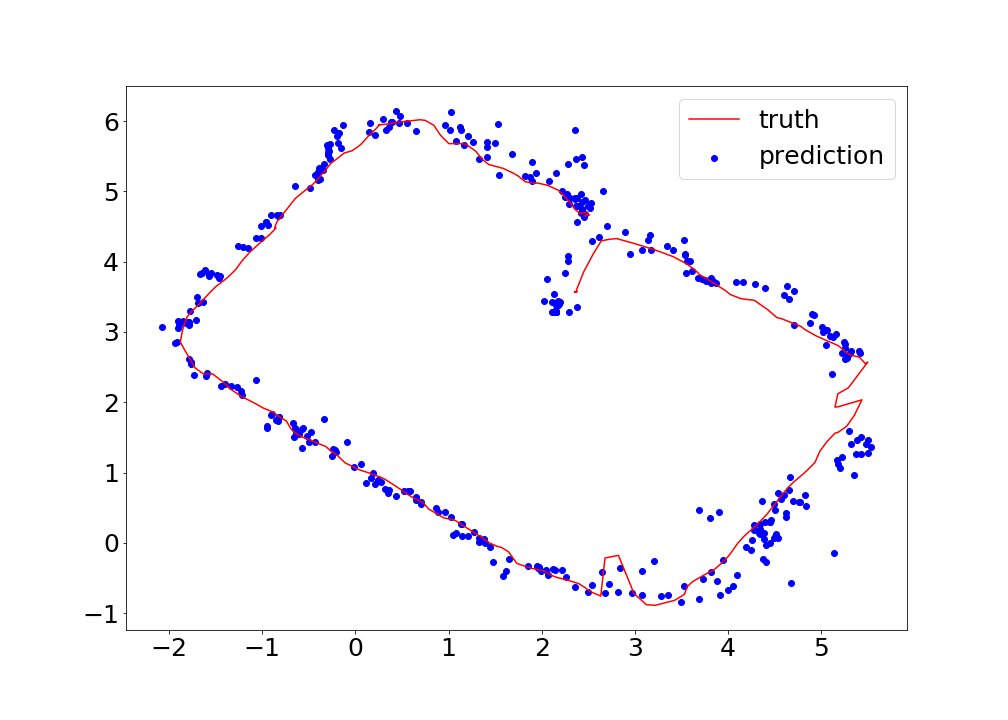}
    \caption{Raw single-point predictions before post-processing.}
    \label{fig:Single}
\end{figure}

\begin{figure}[H]
    \centering
    \includegraphics[width=0.8\linewidth]{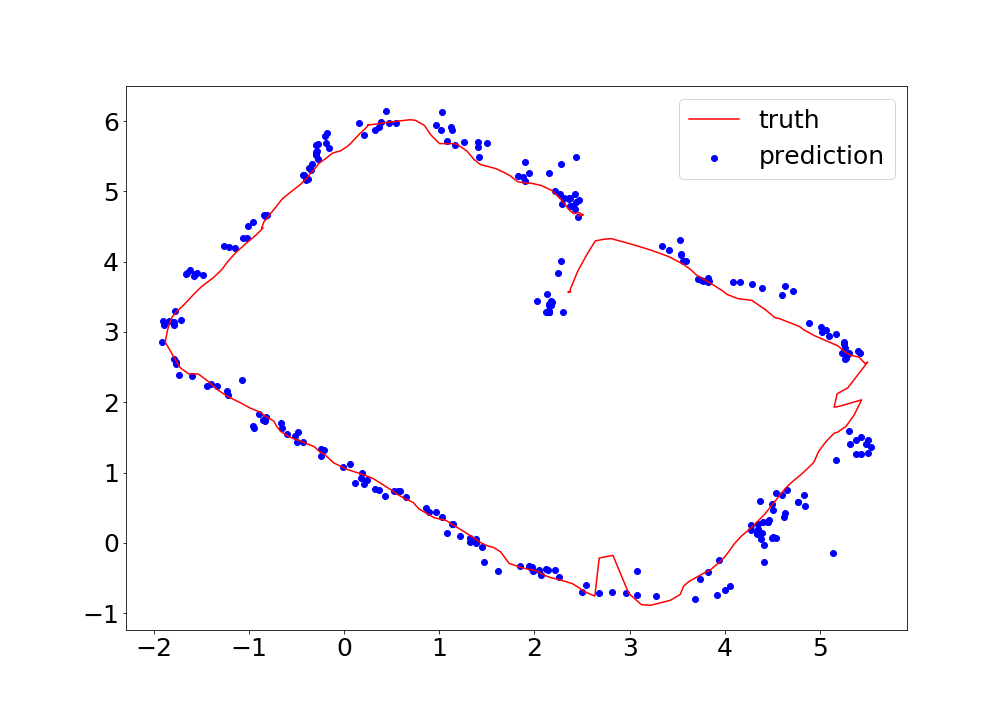}
    \caption{Results after outlier detection and filtering.}
    \label{fig:filter}
\end{figure}

\begin{figure}[H]
    \centering
    \includegraphics[width=0.8\linewidth]{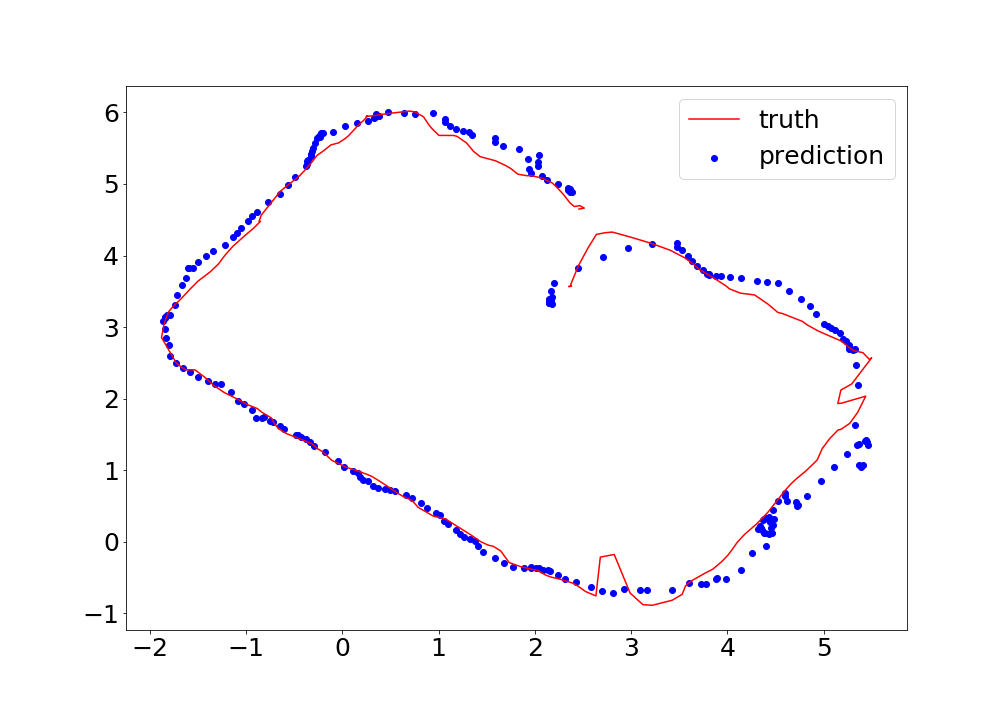}
    \caption{Final trajectory after smoothing and sequence optimization.}
    \label{fig:smooth}
\end{figure}

Figures~\ref{fig:Single}–\ref{fig:smooth} illustrate the impact of the post-processing pipeline on localization accuracy.  
In the raw predictions (Fig.~\ref{fig:Single}), noticeable outliers deviate from the ground-truth trajectory, although most predicted points remain close to the true path.  
After applying outlier filtering (Fig.~\ref{fig:filter}), isolated erroneous points—those lying far from their spatial neighbors—are effectively removed. However, the remaining points still exhibit loose spacing and discontinuities along the trajectory.  
Following the final smoothing stage (Fig.~\ref{fig:smooth}), the predicted path becomes significantly more continuous and aligns closely with the true motion trace.

\begin{algorithm}[H]
\caption{mean-variance outlier detection}
\label{alg:Framwork}
\begin{algorithmic}[1]
\REQUIRE ~~\\
$S = \left\{p_{1},p_{2},...,p_{n}\right\}$,where $p_{i}$ is $i_{th}$ single point in the area.\\
$K$: Number of outliers to filter out.\\
\ENSURE ~~\\
$Outlier = {o_{1},o_{2},...,o_{k}}$
\STATE Given a sequence of single point prediction $S = \left\{p_{1},p_{2},...,p_{n}\right\}$
\STATE for every point $p_{i}$, calculate outlier index $o_{i}= \frac{p_{i}-\mu}{\sigma}$
\STATE  return top k entries in S with largest outlier index.
\end{algorithmic}
\end{algorithm}


\subsection{Results}

Experiments were conducted across multiple environments to evaluate the proposed localization framework under diverse spatial and material conditions:
\begin{itemize}
    \item Meeting room
    \item Social area
    \item Corridor
    \item Outdoor environment
\end{itemize}

In addition, a dedicated 3D experiment was performed in the social area to assess the model’s performance in height estimation. 
Two additional validation rounds were recorded to assess robustness and reproducibility. Validation samples were uniformly distributed across the test space to ensure unbiased evaluation.

The experiments aimed to investigate the following aspects:
\begin{itemize}
    \item Distance coverage between transmitters
    \item Environmental impact (indoor vs. outdoor)
    \item 2D versus 3D localization performance
    \item Influence of surrounding objects
    \item Comparative model performance
\end{itemize}

Unless stated otherwise, the default setup used three transmitters arranged in an equilateral triangle with 4~m spacing between nodes. For indoor–outdoor comparison, two additional transmitters were introduced to expand the sensing area.

\subsubsection{Model Performance}

Table~\ref{tab:model_performance} summarizes the performance of various algorithms. Metrics include mean localization error, standard deviation, and inference time per sample.

\begin{table}[htbp]
\centering
\begin{tabular}{p{1.9cm} p{2.4cm} p{1.4cm} p{0.8cm}}
\toprule
\textbf{Model} & \textbf{Category} & \textbf{Avg. Error (cm)} & \textbf{Std (cm)} \\
\midrule
KNN-Fingerprint & Baseline & 34.6 & 9.52  \\
Triangulation & Baseline & 37.4 & \textbf{6.37}   \\
\midrule
SVM & Machine Learning & 32.8 & 10.36   \\
Random Forest & Machine Learning & \textbf{19.7} & 7.32 \\
AdaBoost & Boosting & 28.2 & 8.93   \\
GBDT & Boosting & 28.7 & 7.78   \\
LSTM & Deep Learning & 47.3 & 12.3   \\
MLP & Deep Learning & 36.7 & 9.27   \\
\bottomrule
\end{tabular}
\caption{Performance comparison across models. Random Forest achieves the best trade-off between accuracy and stability.}
\label{tab:model_performance}
\end{table}

Considering the key selection criteria (accuracy and robustness), the Random Forest regressor demonstrated the most balanced performance and was chosen as the main model for subsequent experiments.

\subsubsection{Effect of Transmitter Distance}

\begin{figure}[]
    \centering
    \includegraphics[width = 0.9\linewidth]{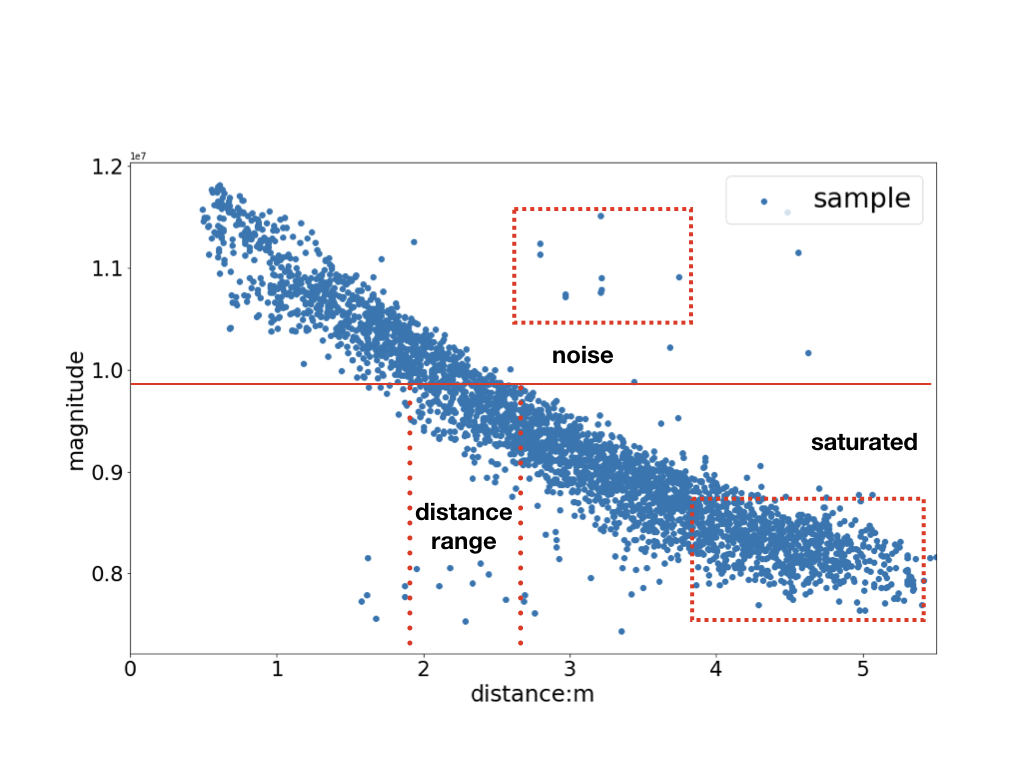}
    \caption{The rectified induced magnetic field magnitude as a function of transmitter–receiver distance, showing nonlinear decay and saturation effects that motivate data-driven modeling.}
    \label{fig:distance-magnitude}
\end{figure}

The transmitter spacing strongly influences both coverage and accuracy. 
The magnetic field gradient decreases with increasing distance, leading to reduced sensitivity beyond 5–6~m. Thus, while shorter distances yield higher accuracy, they cover smaller areas for a fixed number of transmitters.

We tested transmitter spacings of 3~m, 4~m, and 5~m. Results are summarized in Table~\ref{tab:distance_results}.

\begin{table}[htbp]
\centering
\begin{tabular}{cccc}
\toprule
\textbf{Distance (m)} & \textbf{Single-Point (cm)} & \textbf{Sequence (cm)} & \textbf{Std (cm)} \\
\midrule
3 & 18.2 & 13.5 & 5.87 \\
4 & \textbf{19.7} & \textbf{15.4} & 7.32 \\
5 & 30.7 & 22.7 & 9.54 \\
\bottomrule
\end{tabular}
\caption{Localization accuracy at different transmitter distances.}
\label{tab:distance_results}
\end{table}

The 4~m configuration provides the best balance between coverage and accuracy. Beyond 5~m, the magnetic field strength saturates and the mapping between field magnitude and distance becomes unstable, leading to degraded accuracy. Nevertheless, for coarse-grained or classification-oriented tasks, the 5~m setup may still be acceptable due to its broader spatial coverage.

\subsubsection{3D Localization}

For 3D localization, the $z$-axis coordinate was added, with height varying from 20~cm to 150~cm. Experiments were performed in the social area using three transmitters. Data were collected across multiple $X$–$Y$ planes at vertical intervals below 10~cm.

Due to the inherent limitations of the ultrasonic ground-truth system—particularly when the receiver approached the beacon plane—the $z$-axis accuracy was lower than that of the $x$–$y$ axes. Nevertheless, by increasing sampling density within the effective transmission range, the impact of height-related error was mitigated. Overall, we got the reported average error for a single point is 29.6 cm. After the outlier detection and smoothing of the prediction, the average error is 23.4 cm.


\subsubsection{Indoor 2D Localization}

We further evaluated the system’s performance across three indoor environments—social area, corridor, and meeting room—to assess robustness against ambient variations.

\begin{table}[htbp]
\centering
\begin{tabular}{lccc}
\toprule
\textbf{Scene} & \textbf{Single-Point (cm)} & \textbf{Sequence (cm)} & \textbf{Std (cm)} \\
\midrule
Social area & 19.7 & 15.4 & 7.23 \\
Meeting room & 24.6 & 19.5 & 8.56 \\
Corridor & 25.5 & 21.3 & 6.35 \\
\bottomrule
\end{tabular}
\caption{Indoor 2D localization performance across different scenes.}
\label{tab:indoor2D}
\end{table}

Accuracy in the meeting room and corridor was slightly lower than in the social area, primarily due to less optimal transmitter placement and metallic interference. The corridor setup, constrained by wall switches, forced a near-linear transmitter alignment, while the meeting room contained metallic window frames and desks. In contrast, the equilateral triangle configuration used in the social area yielded well-distributed magnetic fields and superior generalization.

\subsubsection{Outdoor 2D Localization}

To test generalization, the same experimental configuration was replicated outdoors. Two independent models were trained: one using indoor data and another using outdoor data. Cross-validation between the two datasets assessed the transferability of learned representations.


\begin{table}[htbp]
\centering
\begin{tabular}{p{2.9cm} p{1.3cm} p{1.3cm} p{1.3cm} }
\toprule
\textbf{Scene} & \textbf{Single-Point (cm)} & \textbf{Sequence (cm)} & \textbf{Cross-Val Error (cm)} \\
\midrule
Indoor (empty) & 19.3 & 15.8 & 4.8 \\
Indoor (with objects) & 24.7 & 20.3 & 5.7 \\
Outdoor & 21.7 & 16.4 & 6.4 \\
\bottomrule
\end{tabular}
\caption{Indoor and outdoor localization performance and cross-validation results.}
\label{tab:outdoor}
\end{table}

The results reveal several key findings:
\begin{itemize}
    \item With identical deployment geometry (five transmitters, 4~m spacing, consistent coil orientation), the system achieved comparable accuracy across indoor and outdoor environments.
    \item The presence of metallic furniture and walls increased measurement noise and reduced precision in indoor environments.
    \item Ultrasound-based ground-truth labeling was less stable in cluttered regions, introducing minor bias in training data.
    \item Despite these challenges, the cross-environment validation (e.g., training the model exclusively on indoor datasets and evaluating it on outdoor data collected under the same transmitter geometry, without retraining) showed less than 7~cm degradation, demonstrating strong model generalization.
\end{itemize}

Overall, the proposed IMF-based system achieved stable sub-30~cm accuracy across diverse scenarios, validating its potential as a calibration-free localization framework for both indoor and outdoor applications.

\section{Conclusion}

This paper presented a machine learning–driven magnetic field–based localization system capable of achieving accurate, real-time positioning in both indoor and outdoor environments. The proposed framework integrates a multi-coil transmitter–receiver setup with data-driven modeling to overcome the nonlinear and environment-dependent nature of induced magnetic fields. Once trained, the system requires no recalibration when deployed in new environments, offering high flexibility and ease of deployment.

Two lightweight post-processing techniques—outlier detection and trajectory smoothing—were introduced to enhance prediction stability. These methods significantly improved localization continuity and robustness while adding negligible computational overhead, making the system suitable for real-time applications. Experimental results demonstrated sub–20~cm accuracy for 2D localization and sub–30~cm accuracy for 3D tracking, validating the system’s precision and generalization across various environments.

Future research will focus on extending the framework’s functionality and adaptability. First, the rich signal characteristics from the receiver coils could be exploited to infer device orientation in addition to position. Second, wearable integration would enable motion tracking and activity recognition using the same sensing principle. Finally, testing the system in larger and more diverse environments would further validate its scalability and robustness.

Overall, this work demonstrates the potential of induced magnetic field sensing combined with machine learning as a practical, calibration-free, and environment-agnostic solution for ubiquitous localization.

\bibliographystyle{IEEEtran}
\bibliography{reference}{}

\end{document}